\documentclass[12pt]{elsarticle}

\IfFileExists{srcltx.sty}{\usepackage[active]{srcltx}}

\usepackage{epsfig}
 \usepackage{verbatim}
 \usepackage{graphicx}

\newcommand{\be}{\begin{equation}}
\newcommand{\ee}{\end{equation}}
\newcommand{\bea}{\begin{eqnarray}}
\newcommand{\eea}{\end{eqnarray}}

\begin{document}
\title{A new interpretation of the gamma-ray observations of distant active galactic nuclei}

\author{Warren Essey$^1$ and Alexander Kusenko$^{1,2}$}

\address{$^1$Department of Physics and Astronomy, University of California, Los
Angeles, CA 90095-1547, USA \\
$^2$Institute for the Physics and Mathematics of the Universe,
University of Tokyo, Kashiwa, Chiba 277-8568, Japan}


\begin{abstract}
Gamma-ray telescopes have reported some surprising observations of multi-TeV photons from distant active galactic nuclei (AGN), which show no significant attenuation due to pair production on either the extragalactic background light (EBL), or the photons near the  source.  We suggest a new interpretation of these observations, which is consistent with both the EBL calculations and the AGN models.  Cosmic rays with energies below 50~EeV, produced by AGN, can cross cosmological distances, interact with EBL relatively close to Earth,  and generate the secondary photons observed by $\gamma$-ray telescopes.   We calculate the spectrum of the secondary photons and find that it agrees with the $\gamma$-ray data.   The delays in the proton arrival times can explain the orphan flares, the lack of time correlations,  and the mismatch of the variability time scales inferred from the multiwavelength observations.  The $\gamma$-ray data are consistent with the detection of the secondary photons, which has important ramifications for gamma-ray astronomy, cosmic ray physics, EBL, and the intergalactic magnetic fields (IGMF). 
\end{abstract}

\maketitle

\newpage


The recent observations of multi-TeV photons from distant blazars \citep{Aharonian:2005gh,2009ApJ...693L.104A,2008AIPC.1085..644C} are surprising because their spectra do not show the expected suppressions due to the interactions with extragalactic background light (EBL).  The observations have led to a debate as to whether one should  reconsider the calculations of EBL \citep{2006ApJ...648..774S,Stecker:2007jq,Stecker:2008fp,Gilmore:2009zb}.  However, even if EBL is thinner and softer, the observations of very high energy (VHE) photons are difficult to reconcile with the expected absorption at or near the source.  For example, the intrinsic photon densities around M87 and 3C279 can be inferred from observations, and these densities appear inconsistent with the unattenuated emission of VHE photons \citep{2008AIPC.1085..644C,2008ApJ...685L..23A}.   Among other proposed explanations, it was suggested that photons may  convert into some hypothetical axion-like particles that convert back into photons in the galactic magnetic fields~\citep{2007PhRvD..76l1301D,Simet:2007sa}, or that Lorentz invariance violation may be the reason for the lack of absorption~\citep{Protheroe:2000hp}.  The orphan flashes, the lack of timing correlations between VHE and optical observations \citep{2009ApJ...695..596H} and the mismatch of time scales for  X-ray and VHE variabilities~\citep{Harris:2009wn}, as well as some unusual energy-dependent time delays~\citep{2008PhLB..668..253M}, pose additional questions.  

These puzzling inconsistencies suggest an alternative interpretation of the observations of distant\footnote{For nearby AGN, the primary photons are expected to dominate the signal, and the EBL absorption features should be seen~\cite{Stecker:1992wi,Stecker:1996ma}.  We concentrate on the AGN that are far enough for the secondary photons to be important.} blazars.  Active galactic nuclei (AGN) are believed to produce ultrahigh-energy cosmic rays.  For energies below Greisen--Zatsepin--Kuzmin (GZK) cutoff (i.e., below 50~EeV), the cosmic rays can propagate cosmological distances without interacting.  The deflection of their trajectories depends on the intergalactic magnetic fields (IGMF), which can be very small along the line of sight (only the upper limits can be derived from observations).  If the IGMF are as small as predicted by the constrained simulations \citep{2004JETPL..79..583D}, the trajectories of  high-energy protons are close to the line of sight until the protons enter our galaxy, where the magnetic fields are much stronger\footnote{These results depend on a number of assumptions, and the cosmic ray data may ultimately be the best probe of IGMF~\citep{Sigl:2003ay,Sigl:2004yk}.}.  For cosmic rays with energies $(4-8)\times 10^{19}$~eV, a correlation of the arrival directions with BL Lacertae objects has been reported~\citep{Tinyakov:2001nr}.  For lower energies, the deflection of protons in the galactic magnetic fields makes the association with an extragalactic source impossible.  However, the secondary photons produced by lower-energy cosmic rays point back to the source because most of them are produced well outside our galaxy.    With small but non-negligible probability the cosmic-ray protons can interact with EBL outside our galaxy, at distances (10--100)~Mpc.  The pions produced in the processes $p\gamma \rightarrow p \pi^{0}$ and $p\gamma \rightarrow n \pi^{+}$ quickly decay and produce photons and electrons with very high energies.  These secondary photons can be observed by the Cherenkov telescopes, and they can create an image of the remote source even if all or most of the primary VHE photons produced by the AGN  are filtered out by their interactions with EBL.  Cosmic ray interactions with EBL should also produce neutrinos~\cite{DeMarco:2005kt}.  

Let us consider the hypothesis that all the photons observed by the gamma-ray telescopes from distant AGN are, in fact, secondary photons from interactions of AGN produced cosmic rays with the background photons.  Production of cosmic rays  is subject to model dependent assumptions, but it is generally believed that the output of a single AGN in cosmic rays with energies above $10^{9}$~GeV is $F_{\rm AGN}= (10^{43}-10^{45}) E^{-1}_9$~erg/s, where $E_9$ is energy in units of $10^9$~GeV~\citep{Berezinsky:2002nc}.
The flux can be beamed with the beaming factor $f_{\rm beam}\sim 10-10^3$, which increases the flux of protons from an AGN with a jet pointing in the direction of Earth (blazar).   Some fraction $f_{\rm int}\sim 10^{-3}$ of the protons can interact with EBL close to  Earth and produce pions, which generate photons and electrons as their decay products.  These photons cascade down to energies below the pair production threshold and generate $f_{\rm mult} \sim 10^3$ photons below the threshold~\citep{2009arXiv0903.3649E}.   This series of interactions  produces the flux 
\begin{eqnarray}
 F_\gamma & \sim & 10^{-12} {\rm cm}^{-2} {\rm s}^{-1} \left( \frac{F_{\rm AGN} [E>10^{7}\, {\rm GeV}] }{ 10^{46} {\rm erg/s}} \right) \nonumber \\ 
& \times & \left( \frac{f_{\rm beam}}{100}\right) \left(  \frac{f_{\rm int}}{10^{-3}} \right) \left(
\frac{f_{\rm mult} }{10^3}\right)  \left(
\frac{ \rm 1~Gpc }{D}\right)^2
,
\end{eqnarray}
which is can be large enough, and which is uncertain enough to be consistent with the fluxes detected by $\gamma$-ray telescopes.  In fitting the $\gamma$-ray data we will use the spectral slope $\propto E^{-2}_p $ predicted by models of cosmic ray acceleration~\citep{Berezinsky:2002nc}, and we will choose the normalization based on the observed $\gamma$-ray flux.  We will not make assumptions regarding the luminosity of a given source in cosmic rays, and we will use the luminosity as a fitting parameter, as long as it is consistent with the overall energetics of  AGN~\citep{Berezinsky:2002nc}.  We note, however, that harder intrinsic blazar spectra than those considered here can be produced by relativistic shock acceleration~\citep{2007ApJ...667L..29S}.

\begin{figure}[ht!]
  \begin{center}
      \includegraphics[width=80mm]{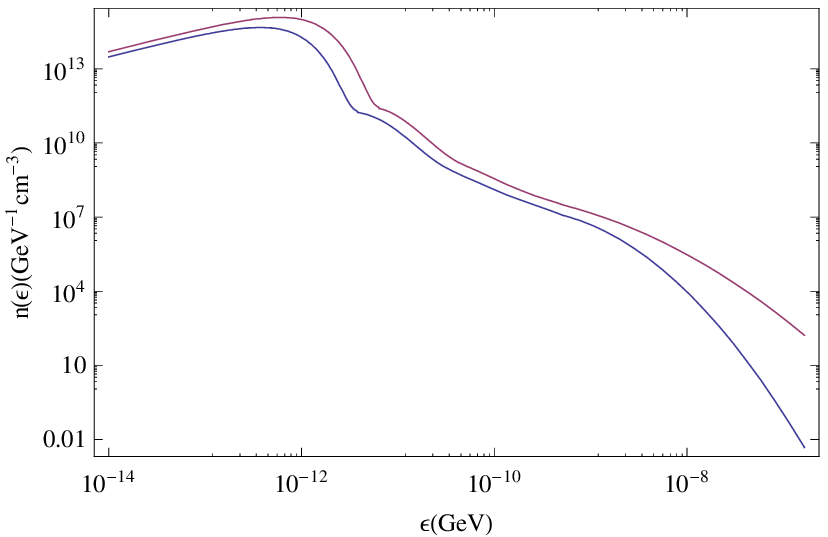}
      \includegraphics[width=80mm]{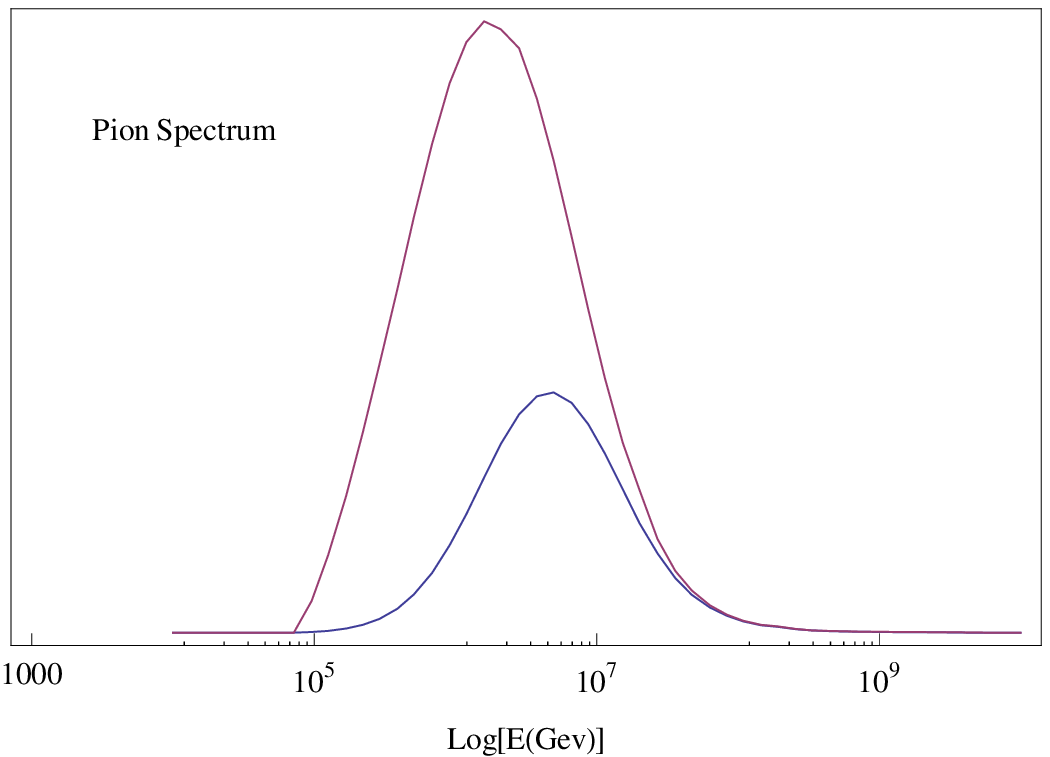} 
   \caption{The mean pion energy, and the corresponding mean photon energy, depend on the spectrum of EBL at a given red shift. 
Left panel shows $n(\epsilon)$ for redshift $z=0$ (lower line) and $z=0.6$ (upper line), according  to Refs.~\cite{2006ApJ...648..774S,Stecker:2007jq}.  The panel on the right shows the spectra of pions produced by $ p\gamma$ interactions at $ z=0.28$ (the spectrum   with a lower mean energy)  and $z=  0.014$ (the spectrum with a higher mean) by protons emitted from the source at $z=0.444$ (e.g., 3C66A) with a $E^{-2}$ spectrum \citep{Berezinsky:2002nc}.  The units are arbitrary, and the upper curve is scaled down by a factor 10. } 
    \end{center}
  \label{ne}
\end{figure}

We have calculated the spectrum of secondary $\gamma$-rays numerically using the EBL spectrum calculated in Refs.~\cite{2006ApJ...648..774S,Stecker:2007jq} (see Fig.~1) in the range $z = 0-0.6$.  The energy losses are due to production of pions in $p\gamma$ interactions with the EBL photons.  The interaction length $\lambda$ is given by \citep{1994APh.....2..375S}
\begin{equation}
\lambda^{-1}=\int^{\epsilon_{max}}_{\epsilon_{min}}\frac{n(z,\epsilon)}{8\epsilon^2E^2}\int^{s_{max}}_{s_{min}}\sigma(s)(s-m_p^2)\, ds \,  d\epsilon, 
\label{lambda}
\end{equation}
where $s_{min}=m_{\pi}^2+m_p^2$, $s_{max}\simeq m_p^2+4\epsilon E_p$ and $\epsilon_{min}\simeq m_{\pi} (m_{\pi}+2m_p)/4E_p$. The photon density $n(z,\epsilon)$ includes both the EBL and cosmic microwave background (CMB) photons, as shown in Fig.~1, and $\epsilon_{max}$ is the cutoff at about 160~eV.  We calculated the optical depth, $\tau(z)$, of the proton traveling from the source at  $z \le 0.6$ to $z=0$ in steps of $\sim $1~Mpc. 
We assume that the protons emitted from AGN have spectrum $\propto E_p^{-2}$ at $10^6-10^9$~GeV, consistent with some acceleration models~\citep{Berezinsky:2002nc}, and we use this spectrum as input for calculating numerically the rates of pion production at various values of the redshift.   These pions, which carry about 12\% of the proton energy,  decay and initiate the electromagnetic cascades~\citep{2009arXiv0903.3649E}. The showers produce both gamma rays and charged particles which should pair produce and undergo inverse Compton scattering (ICS) off the EBL and CMB until their products can propagate the distance to the earth.   The photon flux is integrated numerically over the range of pion production redshifts.

\begin{figure}[ht!]
  \begin{center}
      \includegraphics[width=110mm]{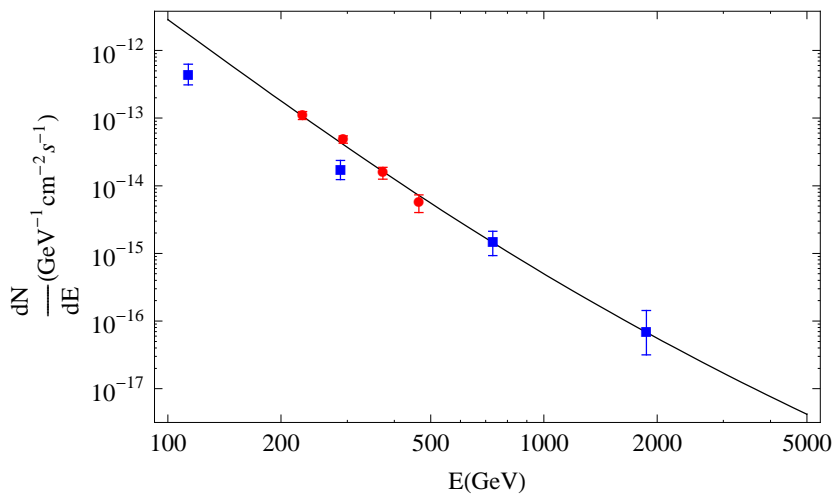}
\caption{Differential spectrum of secondary photons calculated numerically for 3C66A ($z=0.44 $), assuming the proton spectrum at the source $\propto E_p^{-2} $, and normalized to fit MAGIC (blue squares) and VERITAS (red circles) data points. 
}
    \end{center}
  \label{fig:flux}
\end{figure}

The predicted spectrum of $\gamma$-rays turns out to be similar for all the distant AGN.  We have calculated the spectra for redshifts of 3C279, 1ES~1101-232, 3C66A.  The predicted spectra do not show significant differences because the two main factors that affect the spectrum of secondary photons, the intrinsic spectrum of protons and the EBL, are similar for these sources.  We show in Fig.~2  the spectrum of secondary VHE photons that should be observed from 3C66A  ($z=0.44$).  The normalization is chosen to fit the data, and the spectral slope is in reasonable agreement with the observed spectrum of photons.   Assuming the beaming factor $f_{\rm beam}= 10^2$, the best-fit normalization of the flux implies the intrinsic luminosity in cosmic rays $F_{AGN}\sim 3\times 10^{46} {\rm erg/s}$ at energies above $10^7$~GeV.   Of course a higher $f_{beam}$ would imply a lower value of $F_{AGN}$, and vice versa.  The best fit to the spectral slope, $dN/dE_\gamma=f_0E_\gamma^\Gamma$, is $\Gamma=-3.68$ in agreement with MAGIC and VERITAS results. 

Our interpretation of the gamma-ray data is consistent with the observations of point-like sources if the intergalactic magnetic fields (IGMF) are sufficiently small.  Considerations based on Faraday rotation provide some upper limits on the magnetic fields in some directions, but there is no direct observational data for IGMF.  Constrained simulations~\citep{2004JETPL..79..583D,Sigl:2004yk} predict a very inhomogeneous distribution of magnetic fields with voids, filaments, etc., but, as emphasized in Ref.~\cite{2004JETPL..79..583D}, the results of these models should be taken as upper limits on IGMF.   The deflection angles depend on the distance to the source $D$, the correlation length $l_{\rm c}$, and the average value of the magnetic field $B$:
\begin{equation}
 \Delta \theta \sim 0.1^{\circ}
\left( \frac{B}{10^{-14}\rm G}\right)
\left( \frac{4\times 10^{7} {\rm GeV}}{E} \right)
 \left( \frac{D}{1\, \rm Gpc}\right)^{1/2}
\left( \frac{l_{\rm c} }{1\, \rm Mpc}\right)^{1/2}
\end{equation}
It is possible that IGMF along the line of sight are as small as $B\sim 10^{-14}-10^{-13}$~G, in which case the image of a remote blazar  in the secondary photons is point-like.  Even if the line of sight crosses relatively narrow filaments or walls with a higher magnetic field, the angular displacement of the apparent image is not expected to be large, as long as such features are sufficiently distant.  The magnetic fields in the voids can be well below $B\sim 10^{-14}$. 

The magnetic fields of the host galaxy do not affect the pointlike nature of the image.  Even if the proton deflections in the host system are by large angles, $\Delta \theta_{\rm host} \sim 1$, the enlargement of the image cannot exceed $L_{\rm host}/D\ll 0.1^{\circ}$, where $L_{\rm host}$ is the angular size of the host system.  

Of course, the cosmic ray detectors would not observe a point-like image at these energies because of the cosmic ray deflections caused by the galactic magnetic fields, which are much greater than IGMF. 

The photon-rich environments of M87 and 3C279 are not likely to allow the primary TeV photons to come out~\citep{2008AIPC.1085..644C}. However, the cosmic rays are not attenuated significantly by the same backgrounds.  The observations of M87 and 3C279 are consistent with the secondary photon interpretation.  

Multiwavelength observations of blazars~\citep{2009ApJ...695..596H} show the lack of correlation between the optical and VHE bands of variable sources.  Furthermore, recent observations of M87 point to a significant mismatch of variability timescales of the likely emitting region observed in X-rays and in VHE photons~\citep{Harris:2009wn}.  If the VHE photons detected by HESS~\citep{2006Sci...314.1424A}, MAGIC~\citep{2008ApJ...685L..23A}, and VERITAS~\citep{2008ApJ...679..397A} are secondary photons, then both the lack of time correlations and the differences in variability time scales can be explained by the delays in the arrival times of protons as compared to photons.  These delays depend on the IGMF and on the proton energy, $E_p$: 
\begin{equation}
 \Delta t \sim 10^4 {\rm yr} 
\left( \frac{B}{10^{-14}\rm G}\right)^2
\left( \frac{10^{7}\, {\rm GeV}}{E_p} \right)^2
 \left( \frac{D}{1\, \rm Gpc}\right)^2
\left( \frac{l_{\rm c} }{1\, \rm Mpc}\right).
\end{equation}
If IGMF are as low as $10^{-17}$~G, the energy-dependent time delays for the Markarian~501 flares would be of the order of a few minutes.  Since Mrk~501 is relatively close ($z=0.031$), we expect the low-energy signal to be dominated by primary gamma-rays, while the high-energy photons would be secondary photons, arriving with a delay.  This is consistent with the MAGIC observations of energy-dependent delays from a Mrk 501 flare~\citep{2008PhLB..668..253M}.   If this is the case,  no such delay should be seen for more distant sources, such as 3C66A and M87, where all the photons observed should be secondary photons.  The assumption that IGMF are  smaller than $10^{-17}$~G is consistent with both the observational data and some of the models~\cite{Dolag:2004kp}, while other models can accomodate larger IGMF~\cite{Sigl:2003ay,Sigl:2004gi,Sigl:2004yk}. 

The pair 3C66A and 3C66B provides an interesting case.  3C66A is believed to be a blazar at distance 1.7~Gpc.  The radio galaxy 3C66B appears to be only $6^\prime$ away from the line of sight, at distance of 88~Mpc.   VERITAS~\citep{2009ApJ...693L.104A} has reported observations of VHE photons from 3C66A but not 3C66B, while MAGIC has apparently observed photons from 3C66B~\citep{2009ApJ...692L..29A}.  It is possible that the magnetic fields of the radio galaxy close to the line of sight could bend (and focus) the trajectories of the high-energy protons produced by blazar 3C66A.  Therefore, it is possible that MAGIC has observed a flare that originated in 3C66A as a burst of cosmic rays, which were focused by the radio galaxy, and which produced secondary photons pointing back to 3C66B, rather than 3C66A.  

Identification of detected photons with secondary photons from AGN allows one to use the gamma-ray data to study the EBL and the acceleration mechanisms of cosmic rays.  Although the observed spectra depend on the convolution of the two quantities, namely, the EBL and the cosmic ray spectrum,  the future data from similar sources at different red shifts may allow one to deconvolve the uncertainties in these two distributions.

Future studies of gamma-ray data can test our interpretation of the photons that appear to originate in remote AGN.  On the one hand, it is possible that IGMF are high and EBL density is low~\cite{Gilmore:2009zb}, in which case the primary gamma-rays from distant blazars are observed by gamma-ray telescopes.  On the other hand, it is possible that IGMF are low, while EBL density is relatively high~\cite{2006ApJ...648..774S,Stecker:2007jq,Stecker:2008fp}, in which case the most distant AGN are seen almost entirely in secondary photons.  If, indeed, the photons detected on Earth are secondary photons produced by the interactions of cosmic rays with EBL, this association will have a number of ramifications for understanding EBL, IGMF, and the mechanism of cosmic ray acceleration in AGN.  

The authors  thank R.~Ong, F.~Stecker, and V.~Vassiliev for very helpful comments, and M.~Malkan for discussions of EBL and for providing some very useful data.  This work was supported in part  by DOE grant DE-FG03-91ER40662 and by the NASA ATFP grant  NNX08AL48G.

\bibliography{g}
\bibliographystyle{elsarticle-num}

\end{document}